\documentclass[aps,prb,floatfix,amsmath,amssymb,eqsecnum,nofootinbib,twocolumn]{revtex4}
\usepackage{graphicx}% Include figure files
\usepackage{dcolumn}% Align table columns on decimal point
\usepackage{bbm}% bold math
\usepackage{epstopdf}
\usepackage[bookmarks=true,colorlinks,linkcolor=blue,urlcolor=blue,citecolor=blue]{hyperref}
\usepackage{color}

\usepackage{setspace} %line spacing
\newcommand{\beq}{\begin{equation}}
\newcommand{\eeq}{\end{equation}}
\newcommand{\beqn}{\begin{eqnarray}}
\newcommand{\eeqn}{\end{eqnarray}}

\usepackage{amsmath}
\usepackage{amssymb}

\def \q{{\mathbf{q}}}

\def \k{{\mathbf{k}}}

\def \q{{\mathbf{q}}}

\def \x{{\mathbf{x}}}
\def \y{{\mathbf{y}}}

\def \L{{\Lambda}}

\begin{document}
%\preprint{arXiv:1105.xxxx}

\title{Mobile impurity near the superfluid-Mott insulator\\ quantum critical point in two dimensions}

\author{Matthias Punk}
\affiliation{Department of Physics, Harvard University, Cambridge MA
02138}

\author{Subir Sachdev}
\affiliation{Department of Physics, Harvard University, Cambridge MA
02138}

\date{\today }

\begin{abstract}
We consider bosonic atoms in an optical lattice at integer filling, tuned to the superfluid-Mott insulator critical point, and coupled to a single, mobile impurity atom of a different species. This setup is inspired by current experiments with quantum gas microscopes, which 
enable tracking of the impurity motion.
We describe the evolution of the impurity motion from quantum wave packet spread at short times, to Brownian diffusion at long times.
This dynamics is controlled by the interplay between dangerously irrelevant perturbations at the strongly-interacting field
theory describing the superfluid-insulator transition in two spatial dimensions.
\end{abstract}


\maketitle

\section{Introduction}

The problem of a single particle interacting with its environment is encountered frequently in condensed matter physics, and in many variations. 
Such scenarios appear in the form of genuine defects in materials, like magnetic impurities in metals and the associated ubiquitous Kondo problem.\cite{Wilson}
Impurity problems are also often used as a stepping stone to gain insight into the behavior of strongly correlated systems, such as studying a single hole doped into a Mott insulator helps understanding some properties of high-temperature superconductors.\cite{Kane,Vojta,Kaul}

%For example, calculating the properties of a heavy particle strongly coupled to a fermionic bath has been a long standing problem in condensed matter physics, and it took a while to understand to what extent the orthogonality catastrophe survives when the impurity is mobile\cite{RoschKopp}.

In recent years, impurity problems have also started to gain some attention in the context of ultracold atomic gases.
Amongst others, the fermionic polaron problem has been studied thoroughly in connection with strongly imbalanced Fermi gases close to a Feshbach resonance,\cite{Chevy,Combescot,Zwierlein,Grimm} where a single ``spin-down'' atom interacting with a Fermi sea of ``spin-up'' atoms appears as a polaronic quasiparticle and eventually forms a two-particle bound-state with one of the majority atoms if the interactions are strong enough.\cite{Prokofev,Mora,Punk} More generally, such polaronic problems appear when we consider the nature of the threshold of spectral functions of gapped excitations above many-body ground states.\cite{ST,SP}

Here we are interested in a seemingly similar problem, where a mobile impurity is coupled to a continuum of gapless excitations at a quantum critical point, instead of gapless particle-hole excitations at a Fermi surface. In particular, we want to study the situation where bosons in an optical lattice are tuned to the superfluid-Mott insulator transition,\cite{Greiner2002} and a single atom of a different species is coupled to the bosons via a density-density interaction. This scenario is especially interesting due to the recent experimental implementation of quantum-gas microscopes,\cite{Bakr, Kuhr} which allow tracking the motion of the impurity through the bosonic bath. Indeed, a recent combined experimental and theoretical work has addressed a very similar situation in an effectively one-dimensional system.\cite{Bloch} 
We are going to focus on the other experimentally relevant, two dimensional situation instead. From our
perspective, this case also happens to be the most interesting scenario, because in $d=2$ spatial dimensions the critical behavior at the SF-MI transition is described by a non-trivial fixed point without well-defined quasiparticle excitations.\cite{Fisher} Recently there has been a renewed interest in this particular transition as experiments succeeded in observing the so-called Higgs amplitude mode.\cite{Endres, Pollet, Podolsky} 

The main motivation for our work is the prospect of studying aspects of quantum critical transport in a clean, well-defined  experimental setting.
Calculating transport coefficients of systems close to a quantum phase transition remains a challenging problem in theoretical condensed matter physics. The main reason is that all traditional methods for calculating transport properties rely on a quasiparticle picture of the low-energy excitations above the ground-state, which breaks down at a quantum critical point. Indeed, strongly coupled conformal field theories (CFT) that describe physical properties of physical systems close to quantum critical points typically have no well defined quasiparticle excitations.
Some progress has been made recently using the so called AdS/CFT correspondence, which circumvents this problem by mapping the strongly coupled CFT to a dual, weakly coupled gravitational theory, where transport coefficients can be computed reliably.\cite{Herzog}

\begin{figure}
\begin{center}
\includegraphics[width=0.8 \columnwidth]{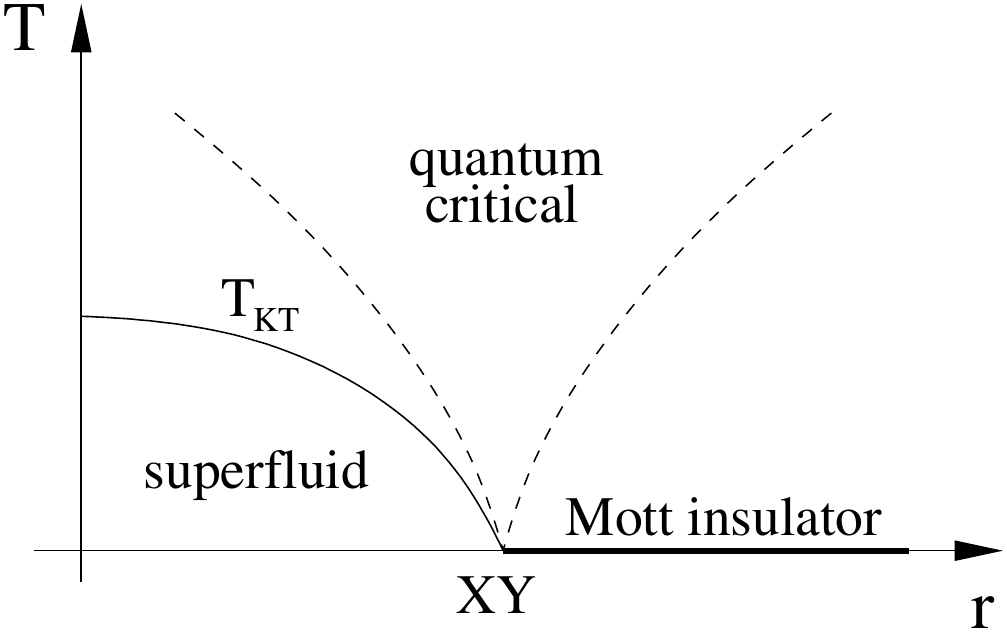}
\caption{Schematic phase diagram of the superfluid - Mott insulator transition of a Bose-Hubbard model in $d=2$ spatial dimensions at integer filling, as a function of temperature $T$ and the tuning parameter $r=\Delta^2-\lambda^2$ (see Eq.~\eqref{lagrangian}). $T_\text{KT}$ marks the Kosterlitz-Thouless transition temperature to the superfluid state. The quantum critical point, marked by XY, is in the universality class of the 3d XY model.\cite{Fisher} We focus on transport properties of an impurity atom coupled to the bosons in the quantum critical regime, indicated by the two dashed crossover lines.}
\label{fig:phase}
\end{center}
\end{figure}

In this paper our main interest is to calculate the diffusion constant $D$ of the impurity as a function of temperature in the quantum critical regime, shown in Fig.~\ref{fig:phase}. This diffusion constant can be measured directly in experiments with quantum-gas microscopes by evaluating the mean square displacement $\langle \x^2 \rangle$ of the impurity atom from its initial position after a time $t$
\begin{equation}
\langle \x^2 \rangle = 4 D t \ ,
\end{equation} 
where the brackets denote an average over many experimental realizations and the time $t \gg \tau$ has to be larger than the typical collision time $\tau$, below which the impurity propagates ballistically. 
Several theoretical works have addressed similar questions using the holographic correspondence in the context of a heavy quark moving through a quark gluon plasma.\cite{karch,teaney,gubser,mukund,sonteaney,chesler,tong}
Here we will use a conventional method based on the Boltzmann equation, however, which turns out to provide an adequate description at low temperatures. The quantum criticality of the bulk critical point has a single characteristic time which controls transport and relaxational
processes close to equilibrium $\tau_R \sim \hbar/(k_B T)$.\cite{Damle} We will show here that the situation with impurity motion
is not as universal, and the impurity time scale is determined by an interplay between several couplings which are formally
irrelevant at the critical point. One of the these couplings is the inverse effective mass, $1/m$, of the mobile impurity; there is the associated
energy scale $mc^2$, where  $c$ the velocity of the bulk bosonic excitations at the critical point. The value of $m$ controls the quantum
wave packet spread of the impurity at short times. 
The other ``irrelevant'' parameters couple the impurity to the bulk excitations, and associated energy scales determine the diffusion
of the impurity at long times even at the lowest temperatures. One of our interesting results is that 
 the diffusion constant of the impurity does \emph{not} depend on $m$ at low temperatures, however.

The rest of the paper is organized as follows. In section \ref{sec:FT} we derive an effective low energy theory for a mobile impurity coupled to bosons at the superfluid--Mott-insulator (SF-MI) transition and we show that the presence of three-body interactions changes the theory significantly at the particle-hole symmetric point. In section \ref{sec:2p} we study perturbative self-energy corrections to the impurity propagator and calculate its diffusion constant as a function of temperature, neglecting three-body interactions. We repeat these calculations including three-body interactions in section  \ref{sec:3p} and show that this changes the results qualitatively at low temperatures. Finally we perform a renormalization group analysis of the effective field theory in Sec.~\ref{sec:RG}.

\section{Field theory approach}
\label{sec:FT}

We start by deriving the effective low-energy theory for a Bose-Hubbard model at the SF-MI critical point in $d=2$ spatial dimensions, following Ref.~\onlinecite{Subir}. A convenient starting point is to consider particle- and hole- excitations on top of a Mott insulating state, which are described by the Lagrangian 
\begin{eqnarray}
\mathcal{L}_c &=& p^* \Big( \partial_\tau +\Delta_p - \frac{\nabla^2}{2 m_p} \Big) p +  h^* \Big( \partial_\tau +\Delta_h - \frac{\nabla^2}{2 m_h} \Big) h \notag \\
&& - \lambda (p^* h^*+ p h) + \dots \ .
\label{Leff}
\end{eqnarray}
Here the fields $p\equiv p(\x,\tau)$ and $h\equiv h(\x,\tau)$ represent particle- and hole excitations on top of a state with an integer filling of bosons. The respective energy gaps for these excitations are denoted by $\Delta_p$ and $\Delta_h$, their effective masses by $m_p$ and $m_h$. The last term in Eq.~\eqref{Leff} creates or annihilates particles and holes in pairs, as required by the conservation of the total number of bosons, and we do not explicitly show higher order terms in the fields $p$ and $h$ that are allowed by symmetry.

Ultimately we are interested in the experimentally relevant situation where the transition happens at a constant density, i.e.~at the tip of the Mott-lobe. This happens only if the gaps $\Delta_p$ and $\Delta_h$ are equal, such that particles and holes condense at the same time when the gap closes. For the moment we will consider the general case, however, and parametrize the gaps as
\begin{eqnarray}
\Delta_p &=& \Delta- \delta/2 \ , \\
\Delta_h &=& \Delta+\delta/2 \ .
\end{eqnarray}
Now we add a single, mobile impurity atom to the bosons, the dynamics of which is described by the free particle Lagrangian
\begin{equation}
\mathcal{L}_\text{imp} = b^* \Big( \partial_\tau - \frac{\nabla^2}{2 m} - \mu \Big) b \ ,
\end{equation}
where the field $b(\x,\tau)$ represents the impurity and the chemical potential $\mu<0$ has to be adjusted such that the density of impurities is zero. This is important only at finite temperatures, however, where we need to ensure that no artificial, thermally excited impurities exist. At zero temperature we can safely set $\mu=0$.

Within a microscopic Bose-Hubbard like model the coupling between the impurity and the bosons is an onsite density-density interaction, which is a pure two-body interaction in the simplest case. Since the local boson density is given by the difference between particle- and hole- densities (up to the constant average number of bosons that we neglect in the following, as it only renormalizes the impurity's chemical potential), the interaction term in our effective field theory takes the form
\begin{equation}
\mathcal{L}_\text{int}^{(2)} = u_2 \,  |b|^2 \, (|p|^2-|h|^2) \ ,
\label{Lint2}
\end{equation}
where the subscript $2$ of the coupling constant $u_2$ indicates that this term derives from a two-body interaction.
Furthermore we include a term that descends from a three-body interaction
\begin{equation}
\mathcal{L}_\text{int}^{(3)} = u_3 \,  |b|^2 \, (|p|^2-|h|^2)^2 \ ,
\label{Lint3}
\end{equation}
because in the case of particle-hole symmetry ($\delta=0$) this interaction generates couplings that are \emph{more} relevant in an RG sense than those deriving from two-body interactions only, as will be shown below. To avoid confusion we note that a three-body interaction in the Bose-Hubbard model of the form $\sim b_i^\dagger b_i \, n_i (n_i-1)$, where $n_i$ denotes the number of bosons on site $i$ and $b^\dagger_i$ is the impurity creation operator, also generates a term of the form \eqref{Lint2} in addition to \eqref{Lint3}. For brevity we will always refer to the term in \eqref{Lint3} as the three-body interaction, however.

Eventually our system is described by the total Lagrangian
\begin{equation}
\mathcal{L} = \mathcal{L}_c + \mathcal{L}_\text{imp} + \mathcal{L}^{(2)}_\text{int} + \mathcal{L}^{(3)}_\text{int} \ .
\label{L0}
\end{equation}
In order to arrive at the more familiar low energy theory for the SF-MI critical point we define the fields
\begin{eqnarray}
\psi &=& (p+h^*)/\sqrt{2} \\
\xi &=& (p-h^*)/\sqrt{2}
\end{eqnarray}
and integrate out the field $\xi$. After a rescaling of $\psi$ and defining $r=\Delta^2-\lambda^2$ as well as $c^2=(\Delta+\lambda) (m_p+m_h)/(4 m_p m_h)$ we arrive at our final result
\begin{eqnarray}
\mathcal{L} &=& \psi^* \Big( \delta \,  \partial_\tau -\partial_\tau^2 - c^2 \nabla^2 + r  \Big) \psi +\frac{g}{4} |\psi|^4 \notag \\
&& + b^* \Big( \partial_\tau - \frac{\nabla^2}{2 m} -\mu \Big) b  + ( \delta  \, u_2 +u_3) \, |b|^2 |\psi|^2 \notag \\
&& + u_2  \big( \psi \, \partial_\tau \psi^* - \psi^* \partial_\tau \psi \big) |b|^2  \ .
\label{lagrangian}
\end{eqnarray}
The terms shown are the leading order terms within a gradient expansion. For the case of particle-hole symmetry ($\delta  = 0$, i.e.~$\Delta_p=\Delta_h$) the dynamical critical exponent of the bosons is $z=1$ and the critical point is described by the well known $U(1)$ CFT in $d+1$ dimensions. Even more interesting, the direct interaction term between the bosons and the impurity $\sim |b|^2 |\psi|^2$ vanishes if only two-body interactions are present. 
This somewhat counterintuitive result (in the sense that the particle-hole symmetric interaction term vanishes if the $\psi^4$-theory is particle-hole symmetric) can be understood from the fact that the two-body interaction term $\mathcal{L}_\text{int}^{(2)}$ is invariant under a particle-hole transformation ($\psi \to \psi^*, \ \delta \to -\delta$) \emph{only} in combination with $ u_2 \to -u_2$. This is not true for the three-body interaction term $\mathcal{L}_\text{int}^{(3)}$, however, and that is why the parametrically smaller $u_3$ generates a term that is more relevant than the one generated by two-body interactions at the particle-hole symmetric point. The leading order terms that are consistent with the symmetry 
\begin{equation}
\mathcal{L}[\psi; \delta,u_2,u_3] = \mathcal{L}[\psi^*; -\delta,-u_2,u_3]
\label{syms}
\end{equation}
are the ones shown in Eq.~\eqref{lagrangian}. 
Also note that the two-body interaction $u_2$ couples the impurity density $|b|^2$ to the time-like component of the bosonic current-density $j_0 =  \psi \, \partial_\tau \psi^* - \psi^* \partial_\tau \psi$.

We stress again that we focus solely on the particle-hole symmetric case ($\delta=0$) in the rest of the paper.
A tree-level scaling analysis of the Lagrangian \eqref{lagrangian} using the exact critical exponents of the bosonic CFT shows that both interactions $u_2$ and $u_3$ are irrelevant in $d=2$ spatial dimensions at the p/h-symmetric point.  Indeed, performing a scaling with $z=1$ gives the scaling dimensions $[\psi] = (d-1+\eta)/2$ and $[b] = d/2$ for the fields, as well as $[ |\psi|^2 ] = d+1-1/\nu$ and $[ j ] = d$ for the composite bosonic density- and current operators per definition. Here $\eta$ and $\nu$ denote the anomalous dimension and the correlation length exponent of the bosonic CFT, which take the values $\eta \simeq 0.038$ and  $\nu \simeq 0.67155$ at the quantum critical point in $d=2$ spatial dimensions.\cite{Vicari}
From these scaling dimensions we get $[u_2] = 1-d$ and $[u_3]=1/\nu-d$, thus both couplings are irrelevant in $d=2$. Also note that the curvature of the impurity dispersion is an irrelevant perturbation, as the scaling dimension of the impurity mass $m$  is $[m]=1$. We will perform a more elaborate RG analysis in Sec.~\ref{sec:RG}.

\section{Two-body interactions}
\label{sec:2p}

In this section we analyze the field theory \eqref{lagrangian} at the particle-hole symmetric point ($\delta=0$), assuming that three body interactions are absent (i.e.~$u_3=0$). The Lagrangian \eqref{lagrangian} thus takes the simplified form
\begin{eqnarray}
\mathcal{L} &=& \psi^* \Big( -\partial_\tau^2 - c^2 \nabla^2 + r  \Big) \psi +\frac{g}{4} |\psi|^4  \label{L2simp} \\
&& + b^* \Big( \partial_\tau - \frac{\nabla^2}{2 m} \Big) b  + u_2  \big( \psi \, \partial_\tau \psi^* - \psi^* \partial_\tau \psi \big) |b|^2  \ . \notag
\end{eqnarray}
Note that we dropped the chemical potential since it is not important at zero temperature, as argued above.

\subsection{Perturbative analysis (T=0)}
\label{sec:pert2}

\begin{figure}
\begin{center}
\includegraphics[width=0.6 \columnwidth]{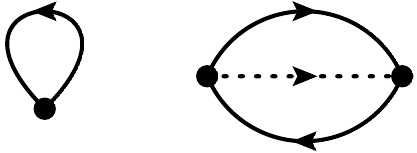}
\caption{Self energy contributions to the impurity propagator up to second order in the interaction $u_2$. Solid lines denote the boson propagator, dashed lines correspond to the impurity propagator.}
\label{fig1}
\end{center}
\end{figure}

As a first step towards an understanding of the impurity's properties we calculate its self-energy perturbatively in the interaction strength $u_2$ at zero temperature in the following. Since $u_2$ is irrelevant, we expect the perturbative calculation to give qualitatively correct results. The terms up to second order are shown in Fig.~\ref{fig1}. The first order tadpole contribution turns out to be identically zero. Note that the boson loop in the second order diagram corresponds to the time-like component of the bosonic current-current correlation function of the free bosonic theory due to the particular form of the interaction vertex in \eqref{L2simp}. At $T=0$ the form of the bosonic current-current correlator of the interacting $\psi^4$-theory at the critical point in $d=2$ is fixed by conformal invariance and is given by
\begin{equation}
K^R_{00}(\q, \omega)= \kappa \, \frac{q^2}{\sqrt{c^2 q^2-(\omega+i 0^+)^2}} \ .
\label{K00}
\end{equation}
Here $\kappa$ is a numerical factor that is renormalized by the boson interaction $g$ and takes the value $\kappa=1/16$ for the free bosonic theory. The functional form of $K_{00}(\q,\omega)$ is independent of $g$, however. For this reason our results for the impurity self-energy below are perturbative in $u_2$, but valid to arbitrary order in the boson interaction $g$.

At zero temperature the imaginary part of the retarded self-energy up to second order in $u_2$ is given by
\begin{equation}
\text{Im} \Sigma_R(\k,\omega) = - u_2^2 \int_\q  \Theta(\omega-\varepsilon_{\k-\q}) \, \text{Im} K^R_{00}(\q, \omega-\varepsilon_{\k-\q}) \ ,
\label{imsig0}
\end{equation}
where $\varepsilon_\k=k^2/(2 m)$ denotes the impurity dispersion. At $\k=0$ the integral can be evaluated exactly and we obtain
\begin{eqnarray}
\text{Im} \Sigma_R(\mathbf{0},\omega) &=& - \frac{u_2^2 \kappa m^2}{\pi} \, \left[ \frac{\omega+m c^2}{2} \log\left(1+\frac{2 \omega}{m c^2} \right)-\omega \right] \notag \\
&=& - \frac{u_2^2 \kappa}{3 \pi c^4} \, \omega^3 + \mathcal{O}(\omega^4)
\end{eqnarray}
\begin{widetext}
For finite momenta we can derive approximate analytic expressions, which take the form
\begin{equation}
\text{Im} \Sigma_R(\k,\delta \omega) \simeq
\begin{cases}
-\dfrac{2 u_2^2 \kappa}{45 \sqrt{2} \, \pi^2} \dfrac{m^4 (29 k^2+31 m^2 c^2)}{(m^2 c^2-k^2)^3} \, \delta \omega^3 + \mathcal{O}(\delta \omega^4)  \  \ & \text{for} \ k<mc  \\ \\
-\dfrac{8 u_2^2 \kappa}{3 \sqrt{2} \, \pi^2} \dfrac{(k-mc)^3}{ c} + \mathcal{O} (\delta \omega, (k-mc)^4) & \text{for} \ k>mc
\end{cases}
\label{pertT0}
\end{equation}
with $\delta \omega = \omega -\varepsilon_\k \geq 0$. 
\end{widetext}
The qualitative change of the self-energy at $k=mc$ can be understood from kinematic considerations. The only lifetime limiting process for the impurity at zero temperature happens to be the excitation of a bosonic "phonon" mode. Due to energy- and momentum conservation this process is prohibited as long as the group velocity of the impurity $v_g = \partial_k \varepsilon_\k=k/m$ is smaller than the "sound" velocity $c$ of the bosons. For this reason the impurity spectral function has a sharp delta-function peak for all momenta $k < m c$, which is reflected by the fact that the imaginary part of the self energy scales as $\sim \delta \omega^3$ close to the quasiparticle pole. On the other hand, for $k> mc$ the impurity can scatter strongly by exciting the bosons. Consequently the imaginary part takes a non-zero on-shell value, implying that there is no well defined impurity excitation anymore. 
The appearance of this kinematic constraint can be readily seen from Eq.~\eqref{imsig0}, where the imaginary part of the current correlator is non-zero only if $\omega > \varepsilon_{\k-\q} + c q$.

It is important to note that this threshold behavior for creating bosonic excitations at $k>mc$ holds to all orders in perturbation theory in the impurity-boson coupling $u_2$. Higher order diagrams with more boson propagators describe processes where the impurity excites multiple boson modes, which in turn have an even higher energetic threshold. From a theoretical perspective it is an interesting question, if the impurity's properties above this kinematic threshold at $k=mc$ are described by a RG fixed point. This is not the case, however, and  we will come back to this problem in Sec.~\ref{sec:RG}.

\subsection{Diffusion constant in the quantum critical and Mott regimes}
\label{sec:Boltz2}

An important transport coefficient that can be directly measured in experiments is the diffusion constant $D$ of the impurity.
Here we calculate $D(T,m,\Delta)$ as a function of temperature $T$, impurity mass $m$ and the Mott gap $\Delta$ using the Boltzmann equation, which gives qualitatively correct results as long as the impurity has a well defined quasiparticle peak. This condition is satisfied in the temperature regime $T \ll m c^2$ (we set $k_B \equiv 1$ from now on), where the kinematic threshold discussed in the previous section doesn't matter. Let us quickly estimate the experimentally relevant temperature regime before proceeding with the Boltzmann calculation. Typical temperatures in experiments\cite{Bakr, Kuhr} in units of the Hubbard interaction $U$ are on the order of $T/U \approx 0.1$. Using the theoretical results for the hopping amplitude $J$ as well as the sound velocity $c$ at the critical point of the 2d Bose-Hubbard model\cite{Capogrosso}, which take the values $(J/U)_\text{crit} = 0.0597$ and $(c/J)_\text{crit} \simeq 4.8$, together with the band mass $m =2/J$ of the impurity, we obtain $T / (m c^2) \approx 0.04$, which is typically well below the kinematic threshold. Note that $mc^2 \approx 46 J$ is a relatively large energy scale.

We are interested in a situation without external forces but with an initial density gradient, thus the Boltzmann equation takes the form
\begin{equation}
\partial_t f_\k(\x,t) + \frac{\k}{m} \cdot \nabla_\x \, f_\k(\x,t) = I[f_\k(\x,t)] \ ,
\label{Boltzequ}
\end{equation}
with $f_\k(\x,t)$ as the Wigner distribution function of the impurity at momentum $\k$, position $\x$ and time $t$. The collision integral has the standard form
\begin{equation}
I[f_\k] = - \sum_{\k'} \big( W_{\k,\k'} f_\k - W_{\k',\k} \, f_{\k'} \big) \ ,
\label{collInt}
\end{equation}
where the transition rates $W_{\k,\k'}$ can be calculated using Fermi's golden rule and are given by 
\begin{eqnarray}
W_{\k,\k'} &=& 2 \, u_2^2 \,  \Big[ \text{Im} K^R_{00}(\k'-\k,\varepsilon_{\k'}-\varepsilon_{\k}) \times \notag \\  \,
&& \times  n_B(\varepsilon_{\k'}-\varepsilon_{\k}) \, \Theta(\varepsilon_{\k'}-\varepsilon_{\k}) \notag \\
&&+ \text{Im} K^R_{00}(\k'-\k,\varepsilon_{\k}-\varepsilon_{\k'})  \times \notag \\ 
&& \times \big( 1+ n_B(\varepsilon_\k - \varepsilon_{\k'}) \big)  \, \Theta(\varepsilon_{\k}-\varepsilon_{\k'})  \Big]  \ .
\label{Wkk}
\end{eqnarray}
Here $K^R_{00}(\q,\omega)$ denotes the time-like component of the retarded bosonic current-current correlation function, $n_B$ is the Bose-Einstein distribution function and $\Theta$ denotes the unit step function. Details of the derivation are given in appendix \ref{app:Boltz}.

In typical experimental situations the impurity is initially localized, released at some time $t=0$ and then is allowed to propagate through the bosonic bath. In principle the Boltzmann equation \eqref{Boltzequ} allows to calculate the full time evolution of the impurity distribution function. At short times collisions don't play a role and the impurity propagates ballistically
\begin{equation}
f_\k(\x,t \ll \tau) = f_\k(\x-\k t / m,0) \ , 
\label{shorttimes}
\end{equation}  
where $\tau$ denotes the typical time between collisions. 
If we start at $t=0$ with the particles localized in a Gaussian wavefunction
\begin{equation}
\psi (\x , 0) = \frac{1}{\sqrt{\pi} \sigma} \exp \left( - \frac{x^2}{2 \sigma^2} \right) 
\end{equation}
then the short time evolution of the Wigner distribution function according to Eq.~\eqref{shorttimes} is free particle behavior
\begin{eqnarray}
f_\k (\x, t \ll \tau) &=& \int d^2 y \, e^{- i \k \cdot \y} \psi^\ast \left(\x - \frac{\y}{2} ,t\right)  \psi \left(\x + \frac{\y}{2},t \right)   \nonumber \\
&=& 4 \exp \left( - \frac{1}{\sigma^2} \left( \x - \frac{\k}{m} t \right)^2 - k^2 \sigma^2 \right). 
\end{eqnarray}

\begin{figure}
\begin{center}
\includegraphics[width=0.95 \columnwidth]{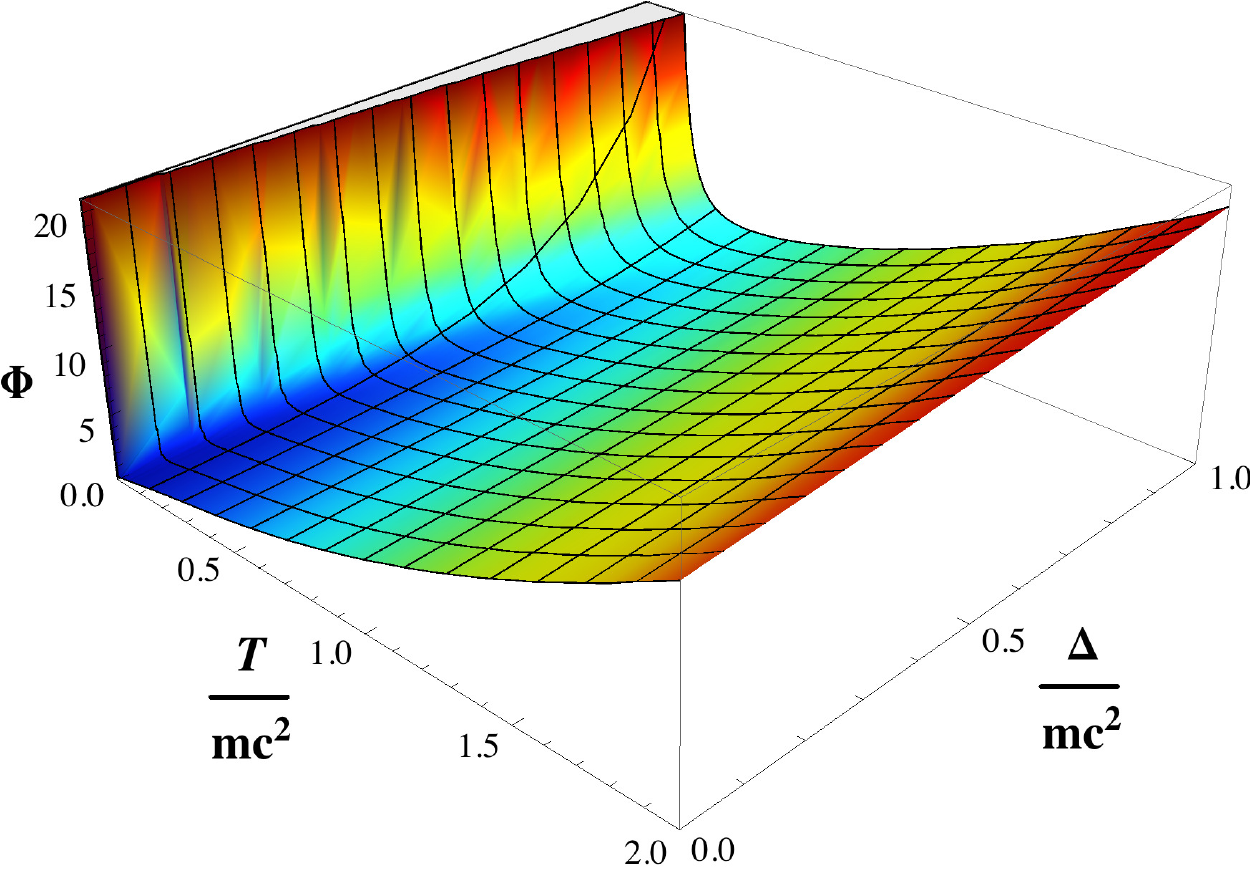}
\caption{(Color online) Scaling function $\Phi\big(\frac{T}{mc^2},\frac{\Delta}{mc^2} \big)$ for the diffusion constant in Eq.~\eqref{Dscale} as a function of temperature $T$ and Mott gap $\Delta$. The scaling function has been evaluated numerically from
Eq.~\eqref{scalingfunc} using the one-loop expression for the current-current correlation function $K_{00}(\q,\omega)$ at finite temperature.}
\label{fig:PhiJ3D}
\end{center}
\end{figure}

At long times $t \gg \tau$ collisions dominate and the impurity propagates diffusively. The Wigner distribution then takes the form 
\begin{equation}
f_\k (\x, t \gg \tau) = \frac{1}{2 m T D t} \exp \left( - \frac{x^2}{4Dt} -  \frac{k^2}{2m T} \right) \ ,
\end{equation}
with a Maxwell-Boltzmann distribution of the momenta.
Indeed, in the long time limit, where the Wigner distribution function of the impurity is close to equilibrium, we can derive a diffusion equation from the Boltzmann equation \eqref{Boltzequ} for arbitrary linear collision integrals of the form \eqref{collInt}. Again, details can be found in appendix~\ref{app:Boltz}. 
The corresponding impurity diffusion constant $D$ takes the simple form
\begin{equation}
D = \frac{T}{m} \, \tau \ , 
\label{diffconst}
\end{equation}
with a transport relaxation time $\tau$ which is given by
\begin{equation}
\tau^{-1} = \sum_{\k,\k'} W_{\k,\k'} \Big[ 1-\frac{k'}{k} \cos \theta_{\k,\k'} \Big] f^\text{Boltz}_\k \ ,
\label{reltime}
\end{equation}
where $k=|\k|$ and $f^\text{Boltz}_\k \sim \exp(-\beta \varepsilon_\k)$ is the equilibrium Boltzmann distribution function. Using the explicit form of the transition rates $W_{\k,\k'}$ in Eq.~\eqref{Wkk} we can cast the expression for the diffusion constant
 $D(T,m,\Delta)$ in a scaling form 
\begin{equation}
D = \frac{c^6}{u_2^2 T^3} \, \Phi \! \left(\frac{T}{m c^2},\frac{\Delta}{m c^2}\right) \ .
\label{Dscale}
\end{equation}
The corresponding scaling function $\Phi$ can be calculated numerically and is shown in Fig.~\ref{fig:PhiJ3D}. Here we used the one loop result for the bosonic current correlator $K_{00}$ at finite temperature, where the dispersion relation of the bosonic modes $E_\k = \sqrt{(c k)^2+m_b^2}$ acquires a temperature dependent mass term  
\begin{equation}
m_b=2 T \log \bigg( \frac{e^{\Delta/(2T)}+\sqrt{4+e^{\Delta/T}}}{2} \bigg) \ ,
\label{massB}
\end{equation}
which follows from a $1/N$ expansion in the limit $N \to \infty$ of the bosonic CFT.\cite{SY} Note that this temperature dependent gap is an artifact of the $N \to \infty$ limit and it is not the scope of this paper to go beyond this limitation. 

\begin{figure}
\begin{center}
\includegraphics[width=0.95 \columnwidth]{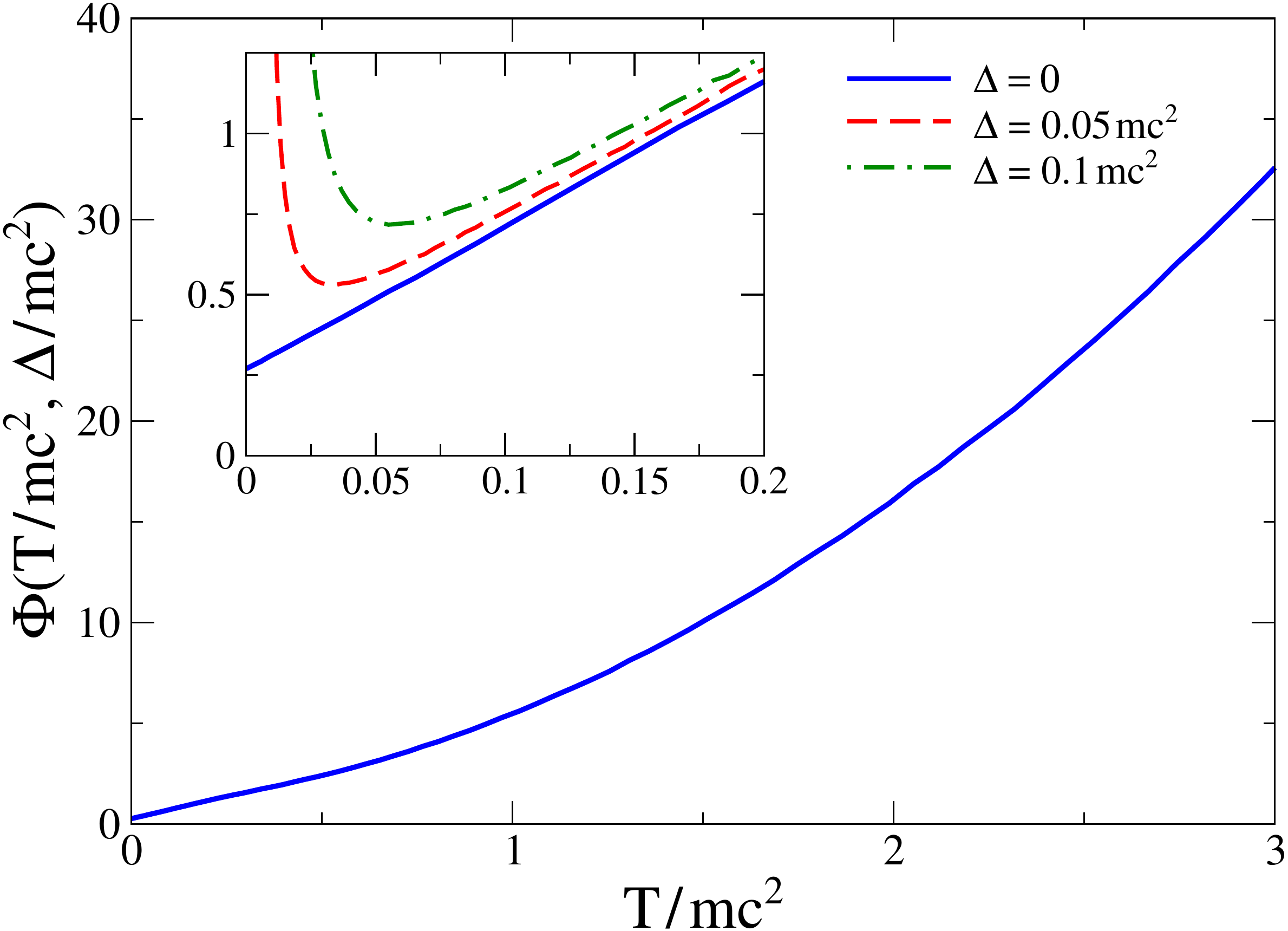}
\caption{(Color online) Scaling function $\Phi\big(\frac{T}{mc^2},0\big)$ for the diffusion constant in Eq.~\eqref{Dscale} as a function of temperature in the quantum critical regime above the critical point, where the Mott gap vanishes ($\Delta=0$). The scaling function has been evaluated numerically from
Eq.~\eqref{scalingfunc} using the one-loop expression for the current-current correlation function $K_{00}(\q,\omega)$ at finite temperature. The inset is an expanded view of the low temperature region, showing that the scaling function $\Phi(x,0)$ is non-zero at $x=0$. The dashed red and dash-dotted green lines show the scaling function in the Mott regime for $\Delta/(m c^2)= 0.05$ and $0.1$ as a function of $T/(m c^2)$, respectively.}
\label{fig:PhiJ}
\end{center}
\end{figure}

At a fixed temperature $T$, the diffusion constant $D$ takes its minimal value in the quantum critical regime directly above the quantum critical point, where the Mott gap vanishes ($\Delta=0$). This is consistent with the naive expectation that the phase space for scattering processes between the impurity and the bosonic modes is large, if the gap of the bosonic modes is small. The scaling function in the quantum critical regime for $\Delta=0$ is shown in more detail in Fig.~\ref{fig:PhiJ}. Note that
$\Phi$ approaches a finite value $\Phi(0,0)  \simeq 0.267$ at zero temperature and scales linearly for small $T/(mc^2) \ll 1$, as
\begin{equation}
\Phi\left(\frac{T}{m c^2},0\right) \simeq 0.267 + 4.4 \,  \frac{T}{m c^2} \ .
\end{equation}
Interestingly, the fact that the scaling function takes a constant value at $T=0$ and $\Delta=0$ implies that the diffusion constant at the critical point does not depend on the impurity mass $m$ at very low temperatures.
Indeed, the curvature of the impurity dispersion, {\em i.e.\/}~the inverse mass $m^{-1}$, is an irrelevant perturbation, and here there is no conspiracy between the two irrelevant couplings $m^{-1}$ and $u_2$ that potentially could change the $T$ dependence of $D$ from
that expected from its $1/u_2^2$ dependence.
%By contrast, the exponent that determines the low temperature dependence of the diffusion constant coincides with the exponent characterizing the low frequency behavior of the self-energy's imaginary part.
Only at higher temperatures the mass starts to play a role, leading to a crossover from a $D \sim T^{-3}$ behavior at very low temperatures to a $D \sim T^{-2}$ dependence at temperatures $T \gtrsim  0.06 \, m c^2$. Note that this crossover temperature is around the typical temperatures in experiments, where $T \gtrsim 0.04 \, m c^2$, thus we expect that the $D \sim T^{-2}$ behavior at larger temperatures should be easily accessible.

In order to compare our results directly to experiments it would be natural to express the diffusion constant in units of the hopping amplitude $J$ (or more precisely $J a^2/\hbar$, if the lattice constant $a$ and Planck's constant are not set to unity). Assuming that the Hubbard interaction $u_2$ between the impurity and the bosons equals the Hubbard-$U$ in the bulk, the diffusion constant at a typical temperature $T=0.1 U$ above the critical point takes the value $D \simeq 3.95 J$. The corresponding transport relaxation time $\tau$ takes the value $\tau \simeq 4.67/J$, showing that it is experimentally promising to reach the diffusive regime at times $t \gg \tau$. 

We note here that in certain parameter regimes a direct comparison of experimental results with our calculations might be difficult due to the harmonic confinement in experiments. This confinement effectively leads to an averaging of the diffusion constant when the impurity explores different regions of the trap. In most cases this is not a problem, however. Only in the Mott regime at temperatures well below the Mott gap, where the diffusion constant is large and the impurity propagates almost ballistically, it explores larger regions of the trap and averaging could become important. Such confinement problems can be circumvented by increasing the temperature slightly, since $D$ and $\tau$ are strongly temperature dependent and decrease by an order of magnitude if the temperature is raised from $T=0.1U$ to $T=0.2U$.

On a formal level, the fact that the scaling function at the critical point $\Phi(0,0)$ is constant can be traced back to the behavior of the current correlator at small frequencies, which scales as $\text{Im}K^R_{00}(\q,\omega) \sim \omega$ for $\omega  \ll T$.
It is important to stress that this behavior, which is related to the fact that $\text{Im}K^R_{00}(\q,\omega)=-\text{Im}K^R_{00}(\q,-\omega)$ is an odd function of frequency and expected to be an analytic at $\omega=0$, is likely valid in general and not related to our one-loop approximation of the current correlator. Indeed it has been argued that the current-current correlation function takes a hydrodynamic form for $\omega,q \ll T$
\begin{equation}
K^R_{00}(\q,\omega) = \chi \, \frac{D q^2}{D q^2- i \omega} \ ,
\end{equation}
the imaginary part of which obviously scales as $\sim \omega$ for small frequencies. This has also been confirmed by calculations using the AdS/CFT correspondence.\cite{Herzog}

In the Mott regime away from the critical point, where the excitation gap $\Delta$ is finite, the scaling function diverges exponentially as the temperature approaches zero
\begin{equation}
\Phi\Big(\frac{T}{mc^2}, \frac{\Delta}{m c^2} \Big) \sim e^{\Delta/T} \ .
\end{equation}
This behavior can be understood from the fact that the impurity does not scatter with bosonic excitations at temperatures well below the Mott gap and thus propagates ballistically, which translates to a diverging diffusion constant.

We also note that the mobility $\mu$ of the impurity is independent of its mass at the critical point at low enough temperatures.
The mobility is defined as $\mu = v_d / F$, where $v_d$ is the impurity's terminal drift velocity in response to an external force $F$. It is related to the diffusion constant $D$ via the Einstein relation $D = \mu T$ and thus has the same dependence on the impurity mass as the diffusion constant.

\subsection{Diffusion constant in the superfluid regime}
\label{sec:BoltzSF}

In the superfluid phase the impurity diffusion constant $D$ can be calculated in the same way as in Sec.~\ref{sec:Boltz2}, using a $1/N$ expansion in the limit $N \to \infty$ for the bosonic current correlation function $K_{00}$. The only difference is that Eq.~\eqref{massB} for the boson mass term has to be replaced by\cite{SY}
\begin{equation}
m_b=2 T \log \bigg( \frac{e^{-2 \pi \Delta_{\text{SF}}/T}+\sqrt{4+e^{-4 \pi \Delta_{\text{SF}}/T}}}{2} \bigg) \ .
\label{mbsuperfluid}
\end{equation}
In contrast to the Mott regime, where the energy scale is set by the Mott gap $\Delta$, the energy scale $\Delta_\text{SF}$ in the superfluid phase is set by the helicity modulus at zero temperature $\Delta_\text{SF}=\rho_s(0)/2$.\cite{SY} 

In the $1/N$ expansion, the helicity modulus is non-zero only at $T=0$, and the O($N$) symmetry is fully restored at any non-zero temperature.
However, for the $N=2$ case of interest here, the helicity modulus is non-zero for a range of low $T$:
the helicity modulus $\rho_s(T)=n_s (T)/M$ is proportional to the superfluid density $n_s(T)$, and 
close to the quantum critical point, $\rho_s(T)$ is a universal function of $\rho_s (0)$ and $T$. 
Nevertheless, we expect the $1/N$ expansion to provide a reasonable description of the impurity atom motion, because the latter couples
only to quantities which are invariant under the U(1) global symmetry of the superfluid. Furthermore, the gap in Eq.~(\ref{mbsuperfluid})
is much smaller than $T$, and so is washed out by thermal excitations. 
In the following we restrict our discussion to the low temperature regime, where the superfluid density is essentially constant and independent of temperature.

\begin{figure}
\begin{center}
\includegraphics[width=0.95 \columnwidth]{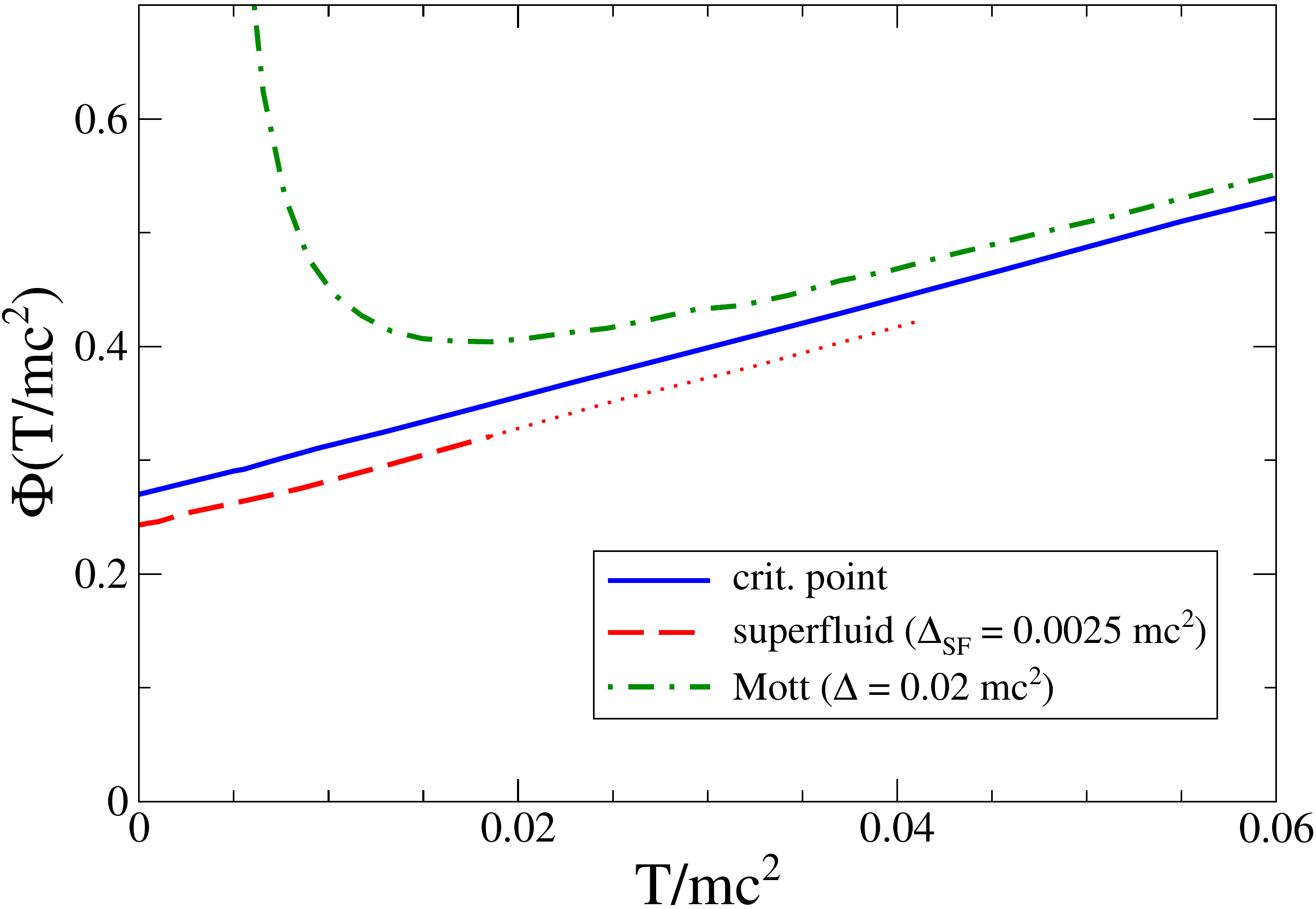}
\caption{(Color online) Scaling function $\Phi\big(\frac{T}{mc^2})$ for the diffusion constant in Eq.~\eqref{Dscale} as a function of temperature in the superfluid regime (shown as red dashed line), together with the results at the critical point (blue solid line) and in the Mott phase (green dash-dotted line). Note that the temperature dependence of $\Delta_\text{SF}$ was neglected in this computation, which is therefore only valid well below the superfluid transition temperature $T_\text{KT}$ (typical values for $T_\text{KT}$ are on the order of the hopping amplitude $J \simeq 0.02 m c^2$). }
\label{fig:PhiJSF}
\end{center}
\end{figure}

Again, we can define a scaling function $\Phi$ for the impurity diffusion constant, which takes the same form as Eq.~\eqref{Dscale}. The scaling function in the superfluid regime is shown in Fig.~\ref{fig:PhiJSF}, together with the results at the critical point and in the Mott regime. Note that the temperature dependence of the diffusion constant in the superfluid phase is qualitatively the same as at the critical point, with the only difference that $D$ takes numerically slightly smaller values in the superfluid phase than at the critical point. This can be understood from that fact the impurity couples efficiently to gapless phonons in the superfluid.

\section{Three-body interactions}
\label{sec:3p}

We discussed in Sec.~\ref{sec:FT}  that three-body interactions are more relevant than two-body interactions at the particle-hole symmetric quantum critical point. 
Indeed, the behavior of the impurity's transport coefficients will be dominated by three-body interactions at low enough temperatures, albeit the three-body interaction being parametrically smaller than the two-body interaction. In this section we thus repeat the analysis of Sec.~\ref{sec:2p} for the case of three-body interactions $u_3$. For the sake of simplicity we set the two-body interaction to zero in this section and start our discussion from the Lagrangian in Eq.~\eqref{lagrangian} with $\delta=0$ and $u_2=0$. In reality both interactions will be present, however, which will lead to a crossover in the temperature dependence of the diffusion constant. The associated crossover scale depends on the relative strength of the two interactions, which will be discussed briefly at the end of this section.

\subsection{Perturbative analysis}

Again, we start by performing a perturbative analysis of the impurity's self-energy at zero temperature, including terms up to second order in $u_3$. The two corresponding self-energy diagrams are analogous to the ones shown in Fig.~\ref{fig1}. Note also that mixed terms $\sim u_2 u_3$ vanish by symmetry.
We neglect the first order tadpole contribution as it only renormalizes the chemical potential, which we set to zero in order to ensure a vanishing density of impurities. The second order contribution to the self-energy can be computed by replacing the current-current correlator $K_{00}(\q,\omega)$ in Eq.~\eqref{imsig0} with the bosonic polarization function $\Pi(\q,\omega)$. Again we want to obtain a result for the impurity self-energy which is non-peturbative in the boson coupling $g$, thus we use the scaling form
\begin{equation}
\Pi_R(\q, \omega) \sim \, \big(c^2 q^2-\omega^2 \big)^{3/2-1/\nu} \ ,
\label{bospolariz}
\end{equation}
for the bosonic polarization function at zero temperature, which follows from the scaling dimensions of the fields discussed in Sec.~\ref{sec:FT}.
The corresponding integral to Eq.~\eqref{imsig0} for the imaginary part of the impurity self-energy can be evaluated exactly at $\k=0$. In the limit of small frequencies $\omega \ll m c^2$ we obtain
\begin{equation}
\text{Im} \Sigma_R(\mathbf{0},\omega) \sim - u_3^2 \,  \frac{ | \sin(\pi(3/2-1/\nu)) |}{4 \pi c^2 (5/2-1/\nu)} \, \omega^{5-2/\nu}
\end{equation}
Using the exact value for the correlation length exponent\cite{Vicari} $\nu=0.67155$, the imaginary part of the self-energy scales as  $\sim \omega^{2.022}$ at low frequencies, i.e.~the impurity spectral function again has a sharp delta-function peak.
Note that this is not true in mean-field theory, where $\nu_\text{MF}=1/2$ and the the imaginary part of the self-energy scales as  $\sim \omega$ at small frequencies, which would result in an algebraic singularity instead of a delta-function peak in the spectral function. 

At finite momenta we expect a similar behavior of $\text{Im} \Sigma$ as described in Sec.~\ref{sec:pert2} for the case of two-body interactions, where the self-energy acquires a finite imaginary part beyond the kinematic threshold at $k = m c$. We calculate this imaginary part explicitly in Sec.~\ref{sec:RG}.

\subsection{Diffusion constant in the quantum critical regime}

For the computation of the impurity's diffusion constant the same arguments hold as in Sec.~\ref{sec:Boltz2} and we restrict our discussion here to the critical point where the Mott gap vanishes ($\Delta=0$).
The transition rates $W_{\k,\k'}$ have the same form as in Eq.~\eqref{Wkk}, with  $u_2^2 \, \text{Im} K^R_{00}(\q,\omega)$ replaced by $u_3^2 \, \text{Im} \Pi_R(\q,\omega) $. 
Using the scaling form of the bosonic polarization function at finite temperature
\begin{equation}
\Pi(\q,\omega,T) =  c^{2/\nu-6} \,   T^{3-2/\nu}\, \tilde{\Pi}(c q/T,\omega/T) \ ,
\end{equation}
we can express the impurity diffusion constant in the scaling form
\begin{equation}
D = \frac{c^{10-2/\nu}}{u_3^2 \, T^{5-2/\nu}} \, \tilde{\Phi}\Big(\frac{T}{m c^2}\Big) \ .
\label{D3}
\end{equation}
Unfortunately we cannot reliably calculate the scaling function $\tilde{\Phi}$ in the equation above, because neither the mean-field approximation with $\nu=1/2$, nor the leading order $1/N$ approximation where $\nu=1$ are good approximations for the polarization function at finite temperature. In fact, from Eq.~\ref{bospolariz} one can see that $\Pi_R(\q,\omega)$ is close to being constant at zero temperature, because the correlation length exponent almost takes the value $\nu \approx 2/3$ at the critical point. However, it is reasonable to assume that the scaling function $\tilde{\Phi}(x)$ in Eq.~\eqref{D3} will again take a constant value at $x=0$, because $\text{Im} \Pi_R(\q,\omega)=-\text{Im} \Pi_R(\q,-\omega)$ is an odd function of frequency and expected to scale linearly $\text{Im} \Pi_R(\q,\omega) \sim \omega$ for small frequencies, similar to the case discussed in Sec.~\ref{sec:Boltz2}.
Based on this assumption the diffusion constant scales as $D \sim T^{-2.022}$ at low temperatures.

If both, two-and three-body interactions are present,  there will be a crossover in the temperature dependence of the diffusion constant. Summing up the two inverse relaxation times using Matthiessen's rule, we obtain
\begin{equation}
D^{-1} \sim u_3^2 \, T^{5-2/\nu} \, \Big(1+ \text{const.} \frac{u_2^2}{u_3^2} \, \frac{T^{2/\nu-2}}{c^{2/\nu-4}} \Big) \ , 
\end{equation}
with a constant of order one. We thus expect a crossover from a $T^{-2.022}$ to a $T^{-3}$ behavior at temperatures $T \gtrsim c \, (u_3/ c u_2)^{\nu/(1-\nu)}$, provided that the mass-dependent crossovers happen at a higher temperature.

\section{RG analysis for the three body-interaction}
\label{sec:RG}

\begin{figure*}
\begin{center}
\includegraphics[width=0.9 \textwidth]{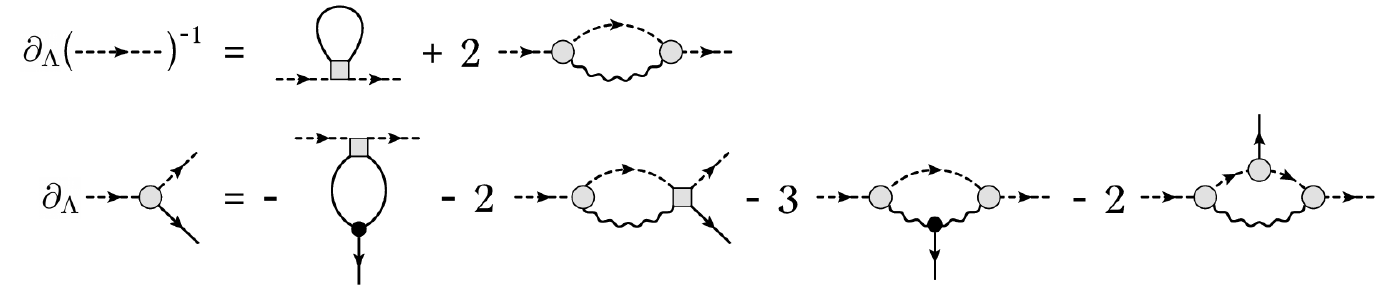}
\caption{Schematic representations of the flow equations for the impurity propagator (dashed line) and the three-point vertex  $u_\L \phi_\L$ (grey circle). Grey squares and black dots denote the four-point vertex $u_\L$ and the boson three-point coupling $g_\L \phi_\L$, respectively.  Insertions of regulator derivatives and associated combinatorial factors are not shown explicitly. The wiggly line is an abbreviation for the difference between the normal and the anomalous boson Green's function ($\mathcal{G}_p-\mathcal{F}_p$). Black lines without arrows denote either $\mathcal{G}_p$ or $\mathcal{F}_p$ and we don't show all combinatorial possibilities explicitly. For details see Eqs.~\eqref{flowequB} and \eqref{flowequU}.}
\label{figFlow}
\end{center}
\end{figure*}

The simple tree level-scaling at the end of Sec.~\ref{sec:FT} showed that both $u_2$ and $u_3$ are irrelevant couplings at the particle-hole symmetric point ($\delta=0$), with $u_3$ being less irrelevant than $u_2$. In this section we perform a more elaborate zero-temperature RG analysis of the three-body interaction $u_3$. Our aim is to understand how a finite imaginary part of the self-energy is generated beyond the kinematic threshold discussed in Sec.~\ref{sec:pert2}, thus we are going to analyze the flow of the impurity propagator at a fixed external momentum. The two-body interaction $u_2$ will be set to zero throughout this section.

We are using a functional RG formulation to derive our flow equations.\cite{Berges} In order to keep the analysis transparent we use simple parametrizations for the propagators and vertices, however. Since the coupling between the impurity and the bosons is a four point coupling and the fRG flow-equations have a one loop structure, we employ the standard trick and consider the flow in the symmetry broken phase in order to generate a non-trivial frequency- and momentum-dependence for the impurity propagator while maintaining a simple parametrization of the interaction vertices.\cite{Strack} This approach is similar to the background field method used in field-theoretic RG approaches.

Our starting point is to choose a simple truncation of the effective action $\Gamma_\Lambda[\psi^*,\psi,b^*,b]$ at momentum scale $\L$ in the symmetry broken phase
\begin{eqnarray}
\Gamma_\Lambda &=& \sum_q \Big[ \psi^*_q \big( -G_{\L,q}^{-1} \big) \psi_q + \frac{g_\Lambda \phi_\Lambda^2}{4} \big(\psi^*_q \psi^*_{-q}+ \psi_q \psi_{-q}\big) \Big] \notag \\
 && + \frac{g_\Lambda \phi_\Lambda}{2} \sum_{q,k} \big( \psi^*_q \psi_{q-k} \psi_k +\text{c.c.} \big) \notag \\
 && + \frac{g_\Lambda}{4} \sum_{q,k,p} \psi^*_k \psi^*_{q-k} \psi_{q-p} \psi_p  + \sum_q b^*_q \big( -B_{\L,q}^{-1}  \big) b_q  \notag \\
&& +  \sum_{q,k,p}  u_\Lambda(\Omega_q) \, \psi^*_{k} b^*_{q-k} b_{q-p} \psi_{p} \notag \\ 
&& + \phi_\Lambda \sum_{q,k}  u_\Lambda(\Omega_q) \left( b^*_q b_{q-k} \psi_k + \text{c.c.} \right)  \ .
\label{Gammaeff}
\end{eqnarray}
Here $\phi_\L$ is the expectation value of the bosonic field $\psi$, the subscript $\Lambda$ denotes scale dependent quantities, we use the shorthand notation $q = (i \Omega_q,\q)$ for sums over bosonic Matsubara frequencies and momenta, and we dropped the subscript on the three-body coupling $u_3 \equiv u$ for notational brevity.
Note that even though $\phi_\L$ is zero at the critical point, the properly rescaled field expectation value $\tilde{\phi}$ will be non-zero and is related to the anomalous dimension of the bosons.

The effective action in Eq.~\eqref{Gammaeff} is a straightforward generalization of Eq.~\eqref{lagrangian} for $\delta=0$ and $u_2=0$ to the symmetry broken phase. The important renormalizations appear in the propagators, which we parametrize as (we set the velocity of bosonic excitations to unity ($c=1$) from now on) 
\begin{eqnarray}
 G_{\L,q}^{-1} &=& - Z_\Lambda^\psi (\Omega_q^2+\q^2) - r_\Lambda - g_\Lambda \phi_\Lambda^2 \\
 B_{\L,q}^{-1}  &=& Z_\L^b i \Omega_q - v_\L \Big(q_x + \frac{\q^2}{2 Q}\Big) + i \gamma_\L \text{sgn} (\Omega_q) \ , \  \ \label{ImpProp}
\end{eqnarray}
where $Z_\L^\psi$ and $Z_\L^b$ are wave-function renormalizations and $\text{sgn}$ denotes the sign function.
Note two important modifications of the impurity propagator $B_{\L,q}$. First, we expanded the dispersion relation around a fixed momentum $\mathbf{Q}=Q \hat{\mathbf{e}}_x$ and its corresponding on-shell frequency $Q^2/(2 m)$ in order to study the flow of the propagator at a fixed external momentum. The group velocity $v=Q/m$ is allowed to flow, which in turn captures renormalizations of the impurity's effective mass. Second, we include a damping term $\gamma$, which will flow to non-zero values above the kinematic threshold. As will become clear later, generating a finite imaginary self-energy $\gamma$ is the only possible way to cure the mass-shell singularity that arises from the large phase space for scattering processes between the (almost) linearly dispersing impurity mode and the linearly dispersing critical bosonic modes above the kinematic threshold.

Also note that we didn't include a chemical potential in the impurity propagator. During all our calculations we enforce a vanishing density of impurities by neglecting the contribution of the impurity propagator's pole (or branch cut, if $\gamma\neq0$) in all diagrams. This procedure is tantamount to adjusting a chemical potential such that the density of impurities is zero.

Apart from the impurity propagator also the three-body coupling $u_\L$ in \eqref{Gammaeff} will acquire an important non-analytic frequency dependence during the flow above the kinematic threshold, which can be parametrized as
\begin{equation}
u_\L(\Omega_q) = u_\L' + i u_\L'' \, \text{sgn} (\Omega_q) \ .
\end{equation}

\begin{figure}
\begin{center}
\includegraphics[width=0.95 \columnwidth]{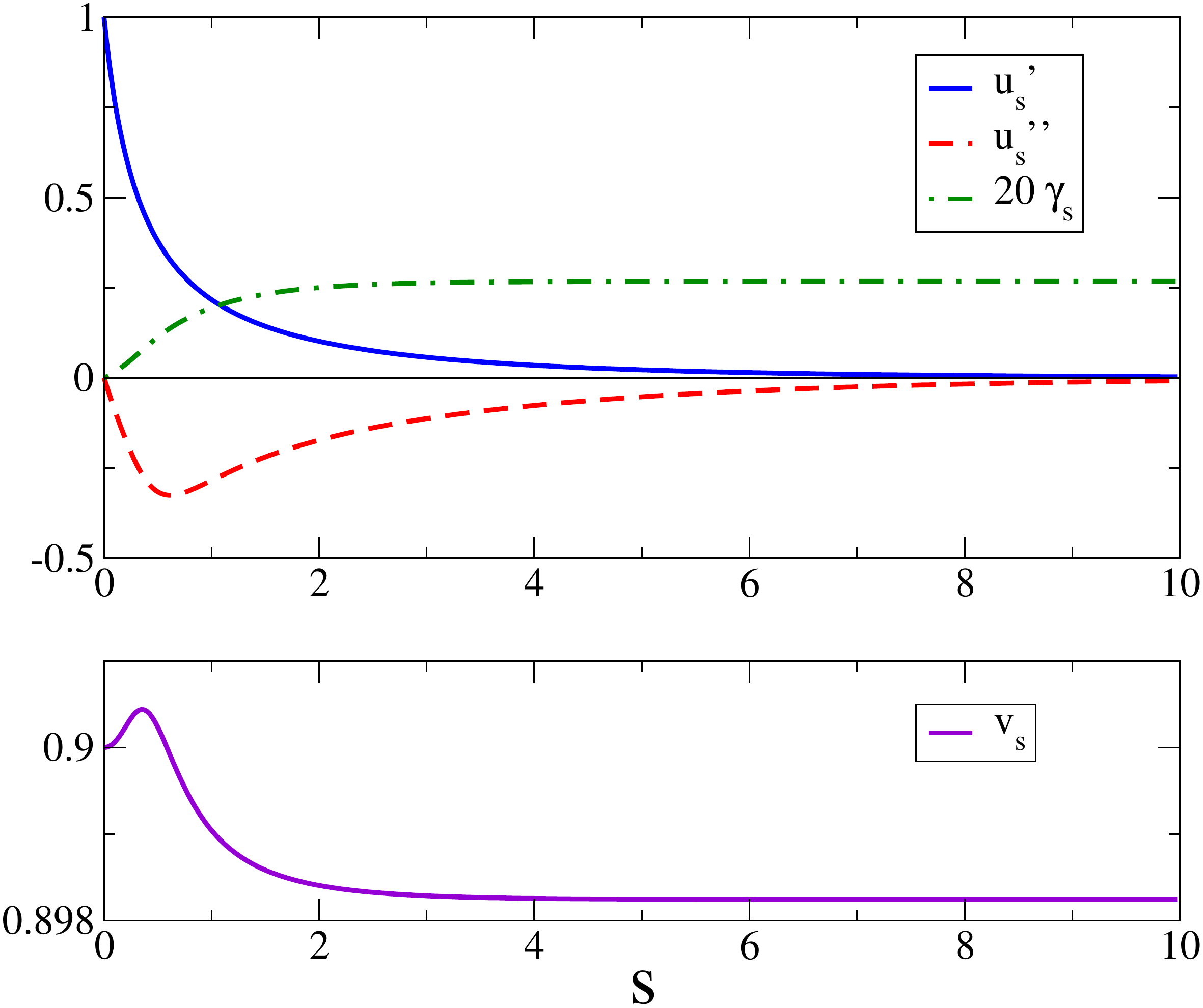}
\caption{(Color online) Upper panel: RG-flow of the rescaled three-body interaction parameters $\tilde{u}'_s$ and $\tilde{u}''_s$ (shown as solid blue and dashed red line) as well as the non-rescaled imaginary part $\gamma_s$ (shown as green dash-dotted line) as a function of the flow parameter $s=\log \L_0/\L$, obtained by integrating the RG equations in appendix \ref{app:A}. The initial conditions were chosen such that the group-velocity $v$ of the impurity is above the kinematic threshold for creating particle-hole excitations. Note that $\gamma_s$ has been multiplied by a factor $20$ for better visibility. Lower panel: corresponding RG-flow of the rescaled impurity velocity $\tilde{v}_s$.}
\label{fig:paramflow}
\end{center}
\end{figure}

Flow equations for all scale dependent quantities can now be derived in the usual way from the exact flow equation for the effective action\cite{Wetterich}
\begin{equation}
\partial_\Lambda \Gamma_\Lambda[\Phi] = \frac{1}{2} \text{Tr} \left[ \frac{1}{\Gamma^{(2)}_\Lambda[\Phi] + R_\Lambda} \, \partial_\Lambda R_\Lambda \right]
\label{Wequ}
\end{equation} 
Here $\Gamma^{(2)}_\Lambda[\Phi]$ is shorthand for the second functional derivative of  the effective action with respect to the fields, i.e.~$(\Gamma^{(2)}_\Lambda[\Phi])_{kq} = \delta^2 \Gamma /(\delta \Phi_{k,\alpha} \delta \Phi_{q,\beta})$ has a $4\times4$ matrix structure coming from the field index $\alpha,\beta=1,...,4$ with $\Phi_k  = (\psi^*_k,\psi_k,b^*_k,b_k)$, besides its usual frequency/momentum-space matrix structure.
The trace in Eq.~\eqref{Wequ} is taken with respect to all momentum, frequency and field indices. $R_\Lambda$ denotes the regulator function 
\begin{equation}
(R_\L)_{k,q} = \text{diag}(R^\psi_{\L,k},R^\psi_{\L,k},R^b_{\L,k},R^b_{\L,k}) \, \delta_{k,q} \ ,
\end{equation}
%A convenient way to derive explicit flow equations from \eqref{Wequ} is to define 
%\begin{equation}
%\Gamma^{(2)}_\Lambda[\psi,b] = -G_\Lambda^{-1} + \Sigma_\Lambda[\psi,b]
%\end{equation}
%where $G_\Lambda^{-1}$ is the inverse, non-interacting ($4\times4$-matrix) propagator. Furthermore defining the regulated propagator ${G^R_\Lambda}^{-1} = G_\Lambda^{-1} - R_\Lambda$, the exact flow equation \eqref{Wequ} can be written as
%\begin{equation}
%\partial_\Lambda \Gamma_\Lambda[\psi,b] = -\frac{1}{2} \text{Tr} \left[ \frac{1}{{G^R_\Lambda}^{-1}-\Sigma_\Lambda[\psi,b]} \, \partial_\Lambda R_\Lambda \right]  \label{Wequ1}
%\end{equation} 
%Flow equations for the scale dependent parameters can now be derived straightforwardly by expanding the denominator in \eqref{Wequ1} as a geometric series, taking appropriate functional derivatives and setting the fields to zero.
and we choose an optimized Litim cutoff for both, the bosons as well as the impurity\cite{Litim}
\begin{eqnarray}
R^\psi_{\L,k} &=& Z_\L^\psi (\L^2-\k^2) \theta(\L^2-\k^2) \ , \\
R^b_{\L,k} &=& \frac{v_\L}{Q} (\L^2-\k^2) \theta(\L^2-\k^2) \ .
\end{eqnarray}
As customary, we neglect derivatives of $Z^\psi_\L$ and $v_\L$ in the scale derivative of the regulator $\partial_\L R_\L$.

The explicit form of our flow equations for all scale-dependent parameters is shown in Appendix \ref{app:A}.
Note that the flow equations for the four boson parameters, $r_\Lambda$, $g_\Lambda$, $\phi_\Lambda$ and $Z_\Lambda^\psi$ are decoupled from the impurity flow equations. This is because we only consider the case of a vanishing impurity density where all diagrams with internal impurity loops evaluate to zero and thus do not give a contribution to the flow equations. For this reason the properties of the bosons at the critical point are described by the usual Wilson-Fisher fixed point with critical exponents in the universality class of the 3D XY-model.

Without discussing the standard flow equations for the bosons much further, we now turn to the impurity flow equations, shown  in Fig.~\ref{figFlow} in a very schematic way. The equations for the impurity propagator and the three-point vertex are given by
\begin{widetext}
\begin{eqnarray}
\partial_\Lambda  B_{\L,q}^{-1} &=& u_\L \sum_p ( \mathcal{G}_p^2+\mathcal{F}_p^2 ) \, \dot{R}^\psi_{\L,p} +2 \, u_\L^2 \phi_\L^2 \sum_p \Big[ (\mathcal{G}_p-\mathcal{F}_p)^2 \mathcal{B}_{q-p} \, \dot{R}^\psi_{\L,p} + (\mathcal{G}_{q-p}-\mathcal{F}_{q-p}) \mathcal{B}_{p}^2 \, \dot{R}^b_{\L,p} \Big]  \ , \label{flowequB} \\
\partial_\L (u_\L \phi_\L) &=& - u_\L g_\L \phi_\L \sum_p \big[ 2 \, \mathcal{G}_p^3 +3 \, \mathcal{G}_p^2 \mathcal{F}_p +6 \, \mathcal{G}_p \mathcal{F}_p^2+\mathcal{F}_p^3  \big]  \, \dot{R}^\psi_{\L,p} \notag \\
&&- 2 \, u_\L^2 \phi_\L \sum_p \Big[ (\mathcal{G}_p-\mathcal{F}_p)^2 \mathcal{B}_{p} \,\dot{R}^\psi_{\L,p} + (\mathcal{G}_{p}-\mathcal{F}_{p}) \mathcal{B}_{p}^2 \, \dot{R}^b_{\L,p} \Big] \notag \\
&&- 3 \, u_\L^2 g_\L \phi_\L^3  \sum_p \Big[2 (\mathcal{G}_p-\mathcal{F}_p)^3 \mathcal{B}_{p} \, \dot{R}^\psi_{\L,p} + (\mathcal{G}_{p}-\mathcal{F}_{p})^2 \mathcal{B}_{p}^2 \, \dot{R}^b_{\L,p} \Big] \notag \\
&&- 2 \, u_\L^3 \phi_\L^3  \sum_p \Big[(\mathcal{G}_p-\mathcal{F}_p)^2 \mathcal{B}_{p}^2 \, \dot{R}^\psi_{\L,p} + 2 (\mathcal{G}_{p}-\mathcal{F}_{p}) \mathcal{B}_{p}^3 \, \dot{R}^b_{\L,p} \Big]  \ , \label{flowequU}
\end{eqnarray}
\end{widetext}
where $\dot{R}=\partial_\L R$ is the scale derivative of the regulator and $\mathcal{G}_p$, $\mathcal{F}_p$  and $\mathcal{B}_p$ denote the regularized normal and anomalous boson Green's functions as well as the regularized impurity Green's function, respectively.
\begin{eqnarray}
\mathcal{G}_p &=& \frac{G_{\L,p}^{-1}-R^\psi_{\L,p}}{(G_{\L,p}^{-1}-R^\psi_{\L,p})^2-(g_\L \phi_\L^2/2)^2} \\
\mathcal{F}_p &=& \frac{g_\L \phi_\L^2/2}{(G_{\L,p}^{-1}-R^\psi_{\L,p})^2-(g_\L \phi_\L^2/2)^2} \\
\mathcal{B}_p^{-1} &=& B_{\L,p}^{-1}-R^b_{\L,p}
\end{eqnarray}

From the explicit form of the flow equations in App.~\ref{app:A} it is clear that the three-body interaction $u$ is always irrelevant. This is mainly due to the screening of the three-body interaction by bosonic particle-hole excitations, giving rise to a contribution to the $\beta$-function of the interaction $u$ which is linear in $u$ and negative, independent of the impurity's properties.

Even though $u$ always flows to zero, a qualitative change in the RG-flow occurs if the group velocity $v$ of the impurity is above the kinematic threshold for creating particle-hole excitations. In perturbation theory this happens for $v>c \equiv 1$, as discussed in Sec.~\ref{sec:pert2}. In our RG procedure the threshold velocity is slightly renormalized to $v_\text{th} = (1+\tilde{r}_*+\tilde{g}_* \tilde{\phi}_*^2/2)^{1/2}$ in units with $c=1$, where $\tilde{r}_*$, $\tilde{g}_*$ and $\tilde{\phi}_*$ denote the fixed-point values of the rescaled variables (see appendix \ref{app:A}). Indeed, for $v>v_\text{th}$ the RG generates a finite damping term $\gamma$ for the impurity. Typical RG-flows of the impurity parameters in the regime $v>v_\text{th}$ are shown in Fig.~\ref{fig:paramflow}. 

\begin{figure}
\begin{center}
\includegraphics[width=0.95 \columnwidth]{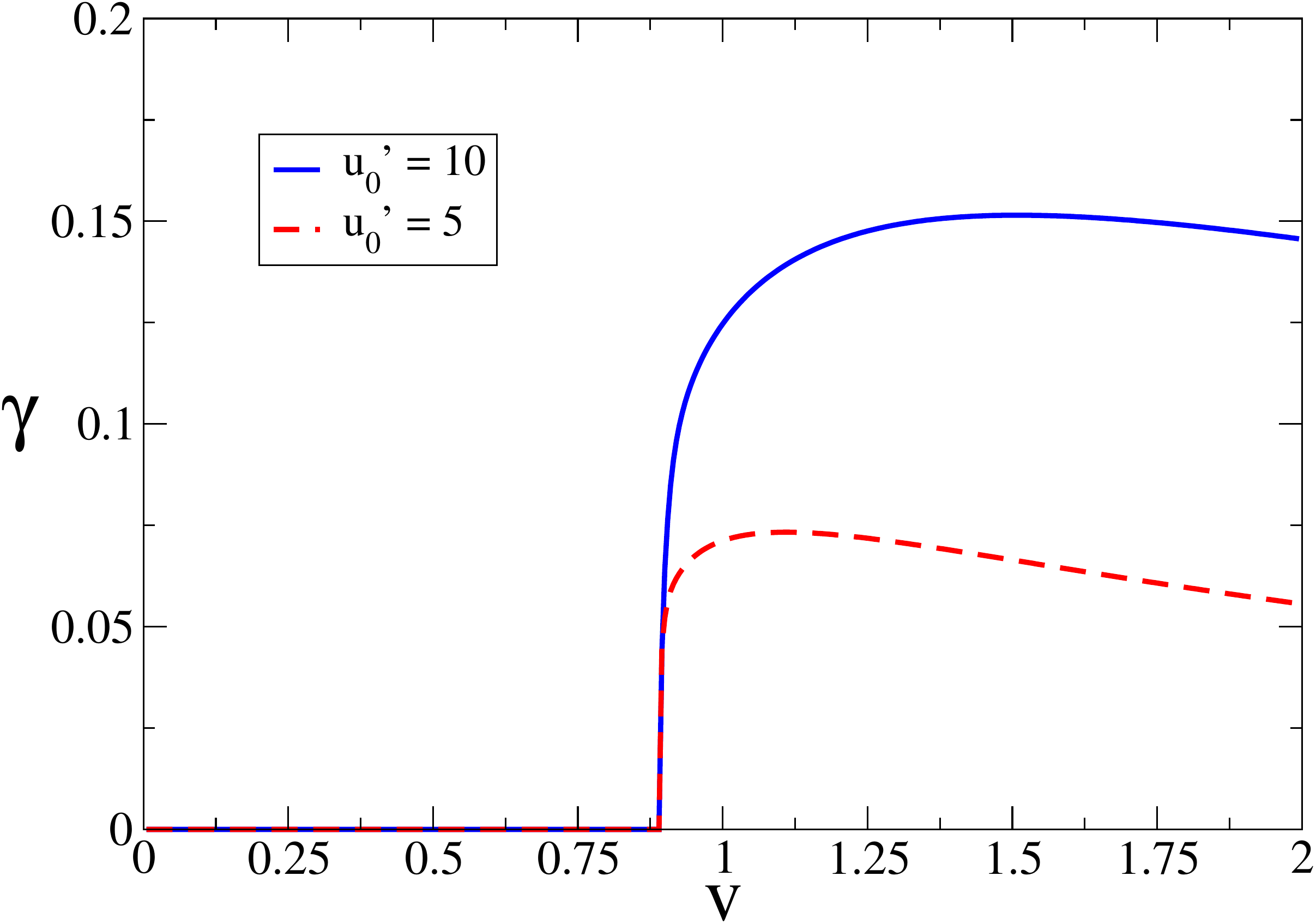}
\caption{(Color online) Imaginary part of the self-energy $\gamma_s$ at the end of the RG-flow ($\L \to 0$) as a function of the impurity's initial group velocity $v_{\L=\L_0}$, plotted for two different initial three-body interaction strengths $u_{\Lambda_0}=10$ (blue solid line) and $u_{\Lambda_0}=5$ (red dashed line). The threshold value of the velocity is renormalized from $v_\text{th}=1$ to $v_\text{th} = (1+\tilde{r}_*+\tilde{g}_* \tilde{\phi}_*^2/2)^{1/2}$ in units where the boson velocity is set to $c=1$ (see text). }
\label{fig:gamma}
\end{center}
\end{figure}

We don't find a RG fixed-point that controls the flow of $\gamma$ above the kinematic threshold. Instead, $\gamma$ flows to a finite, non-universal value while $u$ flows to zero. In Fig.~\ref{fig:gamma} we show the value of $\gamma$ at the end of the flow (i.e.~ at $\L \to 0$) as a function of the initial velocity $v_{\L=\L_0}$, showing the sharp onset of $\gamma$ for $v>v_\text{th}$.
Formally, generating a flow for $\gamma$ is intimately linked to the appearance of a non-integrable mass-shell singularity above the kinematic threshold.

\section{Conclusions}

In this paper we have analyzed the problem of a mobile impurity atom interacting with bosons at the superfluid-Mott insulator critical point in $d=2$ spatial dimensions. The effective low energy theory describing this systems is qualitatively different if three-body interactions are present.
We calculated the diffusion constant $D$ of the impurity as a function of temperature and showed that $D$ does not depend on the impurity mass at low temperatures.
Furthermore we showed that the zero-temperature properties of the impurity change if its group velocity exceeds the kinematic threshold for creating particle-hole excitations. This behavior is not controlled by a RG fixed-point, however. Instead the imaginary part of the impurity's self-energy flows to a non-universal value while the interaction flows to zero.

\acknowledgements

We gratefully acknowledge interesting discussions with I.~Bloch, M.~Endres, T.~Fukuhara, C.~Gross, S.~Hild and in particular P.~Strack.
This research was supported by the National Science Foundation under grant DMR-1103860, 
and by U.S. Army Research Office Award W911NF-12-1-0227.
MP is supported by the Erwin Schr\"odinger Fellowship J 3077-N16 of the Austrian Science Fund (FWF).

\appendix

\section{Derivation of the Boltzmann equation}
\label{app:Boltz}

In this appendix we derive the collision integral for the Boltzmann equation in Sec.~\ref{sec:Boltz2} as well as the diffusion equation, which describes the propagation of the impurity at long times. Our collision integral will be perturbative in $u_2$, but we want to make sure that effects to all order in the boson coupling $g$ are accounted for, similar to the perturbative calculation in Sec.~\ref{sec:pert2}. 
This can be done without invoking the Keldysh formalism by considering an effective model where the impurity is coupled to a bath of phonons. As will be shown in the following, the spectral density of the phonons needs to be chosen such that it matches the spectral function of the bosonic current-current correlation function. The collision integral for the Boltzmann equation can then be derived straightforwardly from the effective phonon model using Fermi's golden rule.

We start from the effective action describing the impurity coupled to a bath of harmonic oscillators
\begin{eqnarray}
S_{\text{eff}} &=& \frac{1}{2} \sum_{k,\alpha} \varphi_{-k,\alpha} (\Omega_k^2+\omega_\alpha^2) \varphi_{k,\alpha} + \sum_k b^*_k(-i \Omega_k+\varepsilon_\k) b_k \notag \\
&&  + \sum_{k,q,\alpha} c_{\k,\alpha} \, \varphi_{k,\alpha} b^*_{q-k} b_q 
\label{Sphon}
\end{eqnarray}
where $\omega_\alpha$ denotes the frequency of the phonon mode $\varphi_{k,\alpha}$, $c_{\k,\alpha}$ are coupling constants and we use the shorthand notation $k=(\Omega_k,\k)$. 
In order to match this effective action to our original action \eqref{L2simp} we integrate out the phonons and obtain
\begin{equation}
S=S_b - \frac{1}{2} \sum_{k,q,q'} \int_0^\infty \frac{d\omega}{\pi} \frac{ J(\k,\omega) \omega}{\Omega_k^2+\omega^2} \, b^*_{q-k} b_q b^*_{q'-k} b_{q'} \ ,
\label{SeffJ}
\end{equation}
where we have defined the spectral density of the phonons via
\begin{equation}
J(\k,\omega) = \pi  \sum_\alpha  \frac{c_{\k,\alpha}^2}{\omega_\alpha} \delta(\omega-\omega_\alpha) \ .
\end{equation}
On the other hand, if we formally integrate out the bosons in our original action \eqref{L2simp} to second order in $u_2$, we get the same result as in Eq.~\ref{SeffJ} with
\begin{equation}
J(\k,\omega) = 2 u_2^2 \, \text{Im} K^R_{00}(\k,\omega) \ .
\end{equation}
Now that we have matched the spectral density of the effective phonon bath to our original model we can straightforwardly employ Fermi's golden rule and obtain the collision integral in Eq.~\eqref{Wkk}.

Close to equilibrium we can solve the Boltzmann equation \eqref{Boltzequ} by first employing a Fourier transform with respect to the spatial coordinate $\x$
\begin{equation}
\partial_t f_\k(\q) - i \frac{\k \cdot \q}{m} \, f_\k(\q) = I[f_\k(\q)] \ ,
\label{BoltzequQ}
\end{equation}
and making the ansatz
\begin{equation}
f_\k(\q) = f^0_k(\q)+(\k \cdot \q) \tilde{f}_k \ .
\label{Boltzansatz}
\end{equation}
Here $f^0_k(\q) \sim n_\q \exp (-\beta \varepsilon_\k)$ and $\tilde{f}_k$ describes the deviation of the momentum distribution from equilibrium.
This is basically a leading order expansion in spherical harmonics where the angular dependence of $f_\k(\q)$ on the momentum $\k$ only enters through the $(\k \cdot \q)$ term. Note that the impurity density distribution is given by
\begin{equation}
n_\q = \sum_\k f_\k(\q) = \sum_\k f^0_k(\q)
\end{equation} 
Using the ansatz \eqref{Boltzansatz} we can derive a diffusion equation from the Boltzmann equation \eqref{BoltzequQ} for arbitrary linear collision integrals of the form \eqref{collInt} as follows.
Taking an angular average of the Boltzmann equation \eqref{BoltzequQ} we obtain
\begin{eqnarray}
\partial_t f^0_k(\q)-i \frac{q^2 k^2}{2 m} \tilde{f}_k &=& \int_{-\pi}^\pi \frac{d \theta_\k}{2 \pi} \, I[f_\k(\q)] \notag \\
&=& I[f^0_k(\q)]  \, \equiv \, 0 \ .
\label{equB1}
\end{eqnarray}
Moreover, taking an angular average over the Boltzmann equation multiplied by $\k \cdot \q$ we get
\begin{equation}
\partial_t^2 f^0_k(\q) + \frac{k^2 q^2}{2 m^2} f^0_k(\q) = \frac{i}{m} \int_{-\pi}^\pi \frac{d \theta_\k}{2 \pi} \, (\k \cdot \q) I[f_\k(\q)] \ ,
\label{equB2}
\end{equation}
where we have used $\eqref{equB1}$ to replace $\tilde{f}$ with $f^0$.
The rhs of the above equation can be evaluated straightforwardly for a collision integral of the form \eqref{collInt}
\begin{eqnarray}
 && \int_{-\pi}^\pi \frac{d \theta_\k}{2 \pi} \, (\k \cdot \q) I[f_\k(\q)] = \label{equB3} \\
&& \ \ \ \  - \frac{q^2 k^2}{2} \sum_{\k'} \big[ W_{\k,\k'} \tilde{f}_k - W_{\k',\k} \frac{k'}{k} \cos \theta_{\k,\k'} \, \tilde{f}_{k'} \big] \ . \notag
\end{eqnarray}
Again, using Eq.~\eqref{equB1} to replace $\tilde{f}$ with $f^0$ in Eq.~\eqref{equB3}, summing over $\k$ and neglecting the $\partial_t^2 f^0$ term, we finally arrive at the diffusion equation
\begin{equation}
\partial_t n_\q = - D q^2 n_\q \ ,
\end{equation}
with the diffusion constant specified in Eqs.~\eqref{diffconst} and \eqref{reltime}.

The explicit expression of the scaling function for the diffusion constant in Eq. \eqref{Dscale} is
\begin{widetext}

\begin{eqnarray}
\Phi^{-1}(x_1, x_2) &=& \frac{1}{2 \pi^2 \, x_1^2} \int_0^\infty dy \int_0^\infty dz \int_{-\pi}^\pi d \theta \, \text{Im} \tilde{K}_{00} \bigg(\sqrt{\frac{2}{ x_1} \, \big( y+2 z-2 \sqrt{z (y+z)} \cos \theta \big)}\, , \ y \, , \  x_2 / x_1  \bigg) \times \notag \\
&& \times \Big[ \big( 1+n(y) \big) \Big(1-\sqrt{\frac{z}{y+z}} \cos \theta \Big) \,  e^{-y} +  n(y) \Big(1 - \sqrt{\frac{y+z}{z}} \cos \theta \Big)  \Big] e^{-z} \ ,
\label{scalingfunc}
\end{eqnarray}
where $\tilde{K}_{00}$ is the scaling function of the current correlator
\begin{equation}
K^R_{00}(\q,\omega, \Delta, T) = \frac{T}{c^2} \, \tilde{K}_{00}(c q/T,\omega/T, \Delta/T) \ ,
\end{equation}
and $n(y)=(e^y-1)^{-1}$. All of the results derived above also apply to the case of three-body interactions if the current correlator $K_{00}(\q,\omega)$ is replaced by the density correlator $\Pi(\q,\omega)$ as well as $u_2$ by $u_3$.

\section{Explicit form of the RG flow equations}
\label{app:A}

It has been shown in Sec.~\ref{sec:RG} that the flow equations for the boson parameters decouple from the impurity flow equations. Indeed, using the rescaled variables
\begin{equation}
\tilde{\phi}_\Lambda = (Z^\psi_\Lambda)^{1/2} \frac{\phi_\Lambda}{\Lambda^{(d-1)/2}} \ , \ \ \ \ \tilde{r}_\Lambda =  \frac{r_\Lambda}{Z^\psi_\Lambda \Lambda^2} \ , \ \ \ \ \tilde{g}_\Lambda =  \frac{g_\Lambda}{(Z^\psi_\Lambda)^2 \Lambda^{3-d}} \\
\end{equation}
and defining the anomalous dimension of the bosons via
\begin{equation}
\eta = - \frac{d \log Z^\psi_\Lambda}{d \log \Lambda}
\end{equation}
the flow equations for the boson parameters are given by ($s=\log \Lambda_0 / \Lambda$)
\begin{eqnarray}
\eta &=& \frac{ \tilde{g}_s^2 \tilde{\phi}_s^2}{4 \pi}  \left( \frac{15}{32 (1+\tilde{r}_s)^{7/2}}  + \mathcal{O}(\tilde{g}_s \tilde{\phi}_s^2) \right)  \label{fequ4} \ , \\
\partial_s \tilde{r}_s &=& (2-\eta) \, \tilde{r}_s + \frac{\tilde{g}_s}{4 \pi} \left( \frac{1}{2 (1+\tilde{r}_s)^{3/2}} +\mathcal{O}(\tilde{g}_s^2 \tilde{\phi}_s^4) \right) \label{fequ1} \ , \\
\partial_s \tilde{\phi}_s  &=& \frac{d-1+\eta}{2} \, \tilde{\phi}_s - \frac{ \tilde{g}_s^2 \tilde{\phi}_s^3}{4 \pi} \left(  \frac{225}{32 (1+\tilde{r}_s)^{7/2}}  + \mathcal{O}(\tilde{g}_s \tilde{\phi}_s^2) \right) \label{fequ2} \ , \\
\partial_s \tilde{g}_s &=& (3-d-2 \eta) \, \tilde{g}_s - \frac{15 \, \tilde{g}_s^2}{32 \pi (1+\tilde{r}_s)^{5/2}}  \left( 1 - \frac{9 \,  \tilde{g}_s \tilde{\phi}_s^2}{1+\tilde{r}_s}  + \mathcal{O}(\tilde{g}_s^2 \tilde{\phi}_s^4)  \right) \ . \label{fequ3} 
\end{eqnarray}
In order to bring the flow equations into an analytically tractable form we expanded them in $\tilde{g}_* \tilde{\phi}_*^2$. This basically amounts to an expansion in a small anomalous dimension, which works well in $d=2$ spatial dimensions. At the Wilson-Fisher fixed point the rescaled variables take the values
\begin{equation}
\eta_* = \frac{1}{29} \approx 0.035 \ , \ \ \ \ \tilde{g}_* \tilde{\phi}_*^{2} = \frac{380}{7749} \approx 0.049 \ , \ \ \ \ \tilde{r}_* = - \frac{28}{123} \approx -0.228
\end{equation}
Note that the value of the anomalous dimension at the critical point $\eta_*=1/29$ agrees pretty well with the expected value for a 3D XY-transition, where $\eta \approx 0.038$.\cite{Vicari}
%The correlation length exponent $\nu$ follows from the largest eigenvalue of the linearized flow equations in the vicinity of the critical point and in our case is given by
%\begin{equation}
%\nu \approx 0.62
%\end{equation}

For the impurity flow equations we define the rescaled variables and the anomalous dimension of the impurity as
\begin{equation}
\tilde{\gamma}_\L = \frac{\gamma_\L}{Z_\L^b \L} \ , \ \ \ \ \tilde{v}_\L= \frac{v_\L}{Z_\L^b} \ , \ \ \ \ \tilde{u}_\L =  \frac{u_\L}{Z_\L^\psi Z_\L^b \L^{2-d}} \ , \ \ \ \ \eta_b = - \frac{d \log Z^b_\Lambda}{d \log \Lambda}
\end{equation}
and the flow equations take the form
\begin{eqnarray}
\eta_b &=& \frac{  \tilde{\phi}_*^{2}}{2 \pi} \text{Re} \Bigg[ (\tilde{u}_s'+i \tilde{u}_s'')^2 \Bigg(\frac{(\alpha-i \tilde{\gamma}_s)^3-i \tilde{\gamma}_s \tilde{v}_s^2}{\tilde{v}_s^2 \alpha^3 \big((\alpha-i \tilde{\gamma}_s)^2-\tilde{v}_s^2 \big)^{3/2}} -\frac{1}{\tilde{v}_s^2 \alpha^3} \Bigg)\Bigg] \ , \\
\partial_s \tilde{u}_s' &=& - \eta_* \tilde{u}_s' - \frac{\tilde{u}_s' \tilde{g}_*}{16 \pi} \Bigg[ \frac{3}{(1+\tilde{r}_*)^{5/2}}-\frac{115  \, \tilde{g}_* \tilde{\phi}_*^{2}}{4 (1+\tilde{r}_*)^{7/2}}\Bigg]-\text{Re} \Bigg[ \frac{(\tilde{u}_s'+i \tilde{u}_s'')^2}{2 \pi} \Bigg(\frac{\tilde{v}_s^2+i \tilde{\gamma}_s (\alpha-i \tilde{\gamma}_s)}{\tilde{v}_s^2 \alpha^3 \big( (\alpha-i \tilde{\gamma}_s)^2-\tilde{v}_s^2\big)^{1/2}} -\frac{i \tilde{\gamma}_s}{\tilde{v}_s^2 \alpha^3} \Bigg) \Bigg] \notag \\
&&+ \frac{3  \tilde{g}_* \tilde{\phi}_*^{2}}{2 \pi} \text{Re}\Bigg[ (\tilde{u}_s'+i \tilde{u}_s'')^2 \Bigg(\frac{3 i \tilde{\gamma}_s (\alpha-i \tilde{\gamma}_s)^3-3 \tilde{v}_s+\tilde{v}_s^2 (4 \alpha^2-9 i \tilde{\gamma}_s \alpha-6 \tilde{\gamma}_s^2)}{4 \tilde{v}_s^2 \alpha^5 \big( (\alpha-i \tilde{\gamma}_s)^2-\tilde{v}_s^2\big)^{3/2}} - \frac{3 i \tilde{\gamma}_s}{4 \tilde{v}_s^2 \alpha^5} \Bigg)\Bigg]  \ ,
\end{eqnarray}
\begin{eqnarray}
\partial_s \tilde{u}_s'' &=& - \eta \tilde{u}_s'' - \frac{\tilde{u}_s'' \tilde{g}_*}{16 \pi} \Bigg[ \frac{3}{(1+\tilde{r}_*)^{5/2}}-\frac{115 \, \tilde{g}_* \tilde{\phi}_*^{2}}{4 (1+\tilde{r}_*)^{7/2}}\Bigg]-\text{Im} \Bigg[ \frac{(\tilde{u}_s'+i \tilde{u}_s'')^2}{2 \pi} \Bigg(\frac{\tilde{v}_s^2+i \tilde{\gamma}_s (\alpha-i \tilde{\gamma}_s)}{\tilde{v}_s^2 \alpha^3 \big( (\alpha-i \tilde{\gamma}_s)^2-\tilde{v}_s^2\big)^{1/2}} -\frac{i \tilde{\gamma}_s}{\tilde{v}_s^2 \alpha^3} \Bigg) \Bigg] \notag \\
&&+ \frac{3  \tilde{g}_* \tilde{\phi}_*^{2}}{2 \pi} \text{Im}\Bigg[ (\tilde{u}_s'+i \tilde{u}_s'')^2 \Bigg(\frac{3 i \tilde{\gamma}_s (\alpha-i \tilde{\gamma}_s)^3-3 \tilde{v}_s+\tilde{v}_s^2 (4 \alpha^2-9 i \tilde{\gamma}_s \alpha-6 \tilde{\gamma}_s^2)}{4 \tilde{v}_s^2 \alpha^5 \big( (\alpha-i \tilde{\gamma}_s)^2-\tilde{v}_s^2\big)^{3/2}} - \frac{3 i \tilde{\gamma}_s}{4 \tilde{v}_s^2 \alpha^5} \Bigg)\Bigg]  \ , \\
\partial_s \tilde{v}_s &=& -\eta_b \tilde{v}_s+\frac{ \tilde{\phi}_*^{2}}{2 \pi} \text{Re} \Bigg[(\tilde{u}_s'+\tilde{u}_s'')^2 \Bigg( \frac{\tilde{v}_s (\alpha-i \tilde{\gamma}_s)^3+i \tilde{\gamma}_s \tilde{v}_s^3}{\tilde{v}_s^2 \alpha^3 \big( (\alpha-i \tilde{\gamma}_s)^2-\tilde{v}_s^2\big)^{3/2}}-\frac{1}{v \alpha^3} \Bigg)\Bigg] \ , \\
\partial_s \tilde{\gamma}_s &=& (1-\eta_b) \tilde{\gamma}_s - \frac{\tilde{u}_s''}{8 \pi} \, \frac{2 (1+\tilde{r}_*)-3 \, \tilde{g}_* \tilde{\phi}_*^{2}}{2 (1+\tilde{r}_*)^{5/2}}+ \frac{\tilde{\phi}_*^{2}}{4 \pi} \text{Im} \Bigg[ (\tilde{u}_s'+i \tilde{u}_s'')^2\Bigg( \frac{\tilde{v}_s^2+i \tilde{\gamma}_s (\alpha-i \tilde{\gamma}_s)}{\tilde{v}_s^2 \alpha^3 \big( (\alpha-i \tilde{\gamma}_s)^2-\tilde{v}_s^2\big)^{1/2}} -\frac{i \tilde{\gamma}_s}{\tilde{v}_s^2 \alpha^3}  \Bigg) \Bigg] \ ,
\end{eqnarray}
where we have defined $\alpha^2 = 1+\tilde{r}_*+\tilde{g}_* \tilde{\phi}_*^2/2$ and fixed all boson parameters at their respective critical values. Again we have expanded boson loops in $\tilde{g}_* \tilde{\phi}_*^2$ and moreover we have taken $Q \to \infty$ for the sake of simplicity, which amounts to neglecting the small curvature of the impurity dispersion in \eqref{ImpProp}. This does not change the results qualitatively.
We also mention that the term $\sim \eta_b u_s$ cancelled exactly with the $\sim u_s^3$ contribution in the flow equation for the three-body interaction $u_\L$. 

The particular structure with real- and imaginary-parts in the impurity flow equations originates from the non-analytic term $\sim i\gamma \, \text{sgn} \Omega_q$ in the boson propagator \eqref{ImpProp}. After setting the external boson frequencies to zero, all diagrams with internal impurity lines depend on the external impurity frequency $\Omega_q$ via the damping term $i \gamma \, \text{sgn} \Omega_q$.  We can extract the contributions of analytic and non-analytic terms by taking limits $\Omega_q \to 0^\pm$ from above and below and adding or subtracting the resulting expressions, which gives rise to the particular form of the impurity flow equations above.

\end{widetext}

\end{document}